\documentclass{ws-procs975x65}

\def\beq{\begin{equation}}
\def\eeq{\end{equation}}

\begin{document}

\title{Cosmology of the de Sitter Horndeski models}
\author{$^1$Nelson J.~Nunes, $^2$Prado Mart\'{\i}n-Moruno and $^1$Francisco S.~N.~Lobo}

\address{$^1$Instituto de Astrof\'isica e Ci\^encias do Espa\c{c}o, Universidade de Lisboa, \\
Faculdade de Ci\^encias, Campo Grande,\\ PT1749-016 Lisboa, Portugal. \\
$^2$Departamento de F\'{\i}sica Te\'orica I, Universidad Complutense de Madrid,\\ E-28040 Madrid,
Spain.
}

\begin{abstract}
Horndeski models with a de Sitter critical point for any kind of 
material content may provide a mechanism to alleviate the cosmological 
constant problem. We study the cosmological evolution of two classes of 
families - the linear models and the non-linear models with shift 
symmetry. We conclude that the latter models can deliver a background 
dynamics compatible with the latest observational data.
\end{abstract}


\bodymatter

\section{Introduction}

The discovery, in 1998, that the Universe is currently undergoing an 
accelerated expansion is one of the greatest milestones in all physics. 
Naturally, over the last 17 years, many proposals to explain this 
evolution have been brought forward. Most ideas involve scalar field 
dark energy or extensions of Einstein's gravity. These proposals are 
essentially phenomenological without any relation to each other. One 
major step forward was the realization in 2011 that all these proposals are 
subclasses of the most general scalar-tensor theory that leads to second 
order equations of motion, the Horndeski Lagrangian.  \cite{Horndeski:1974wa,Deffayet:2011gz}

Seeking viable cosmological solutions, one can focus on the 
Friedmann-Lemaitre-Robertson-Walker spacetime and search for cosmological 
models that have a late time flat de Sitter critical point for any kind of material content or 
value of the vacuum energy.  Such models were attained \cite{modelo}  and in this 
proceedings  we address their phenomenology following recent work. \cite{lineal, nolineal}

\section{Linear models}

It was recently shown \cite{modelo} that it is possible to construct a scalar field Lagrangian, $L(\phi,\dot\phi)$ linear in $\dot \phi$,
such that: (i) the field equation for the Hubble rate, $H(t)$, is trivially satisfied at the critical
point to allow the  field to self-adjust; (ii) at the critical point, the Hamiltonian depends on $\dot\phi$, so that the continuous field can absorb discontinuities of the 
vacuum energy and;  (iii) the scalar field equation of motion depends on $\dot H$, such 
that the cosmological evolution is non-trivial before screening
takes place.  Considering minimally coupled matter, the linear Lagrangian becomes
\begin{equation}\label{L}
L=L_{\rm EH}+L_{\rm linear}+L_{\rm m}, 
\end{equation}
where 
\begin{equation}\label{Ll}
 L_{\rm linear} =  a^3 \sum_i \left(3 \sqrt{\Lambda} U_i(\phi) + \dot\phi W_i(
\phi) \right) H^i,
\end{equation}
with $i = 0,...,3$, subject to the constraint, 
\begin{equation}
 \sum_i W_i(\phi)\Lambda^{i/2}=\sum_j U_{j,\phi}(\phi)\Lambda^{j/2},
 \end{equation}
which ensures that the Lagrangian density evaluated at the critical point has the form
required to allow the field to self-tune. 
In Eq.~(\ref{L}) we have written explicitily an Einstein--Hilbert term contained in $L_{\rm linear}$.
The functions $W_i$ and $U_i$ are related to the $\kappa_j$ functions of the Horndeski Lagrangian and $G_j$ functions of the Deffayet {\it et al.}~Lagrangian. 
As there are a total of eight functions and only one constraint, there are effectively only seven free functions which we coined ``the magnificent seven''.\cite{ensayo,Martin-Moruno:2015hia} The field equation for $H'$ and the Friedmann equation read,
\begin{eqnarray*}
\label{H'}
 H'&=&3\frac{\sum_iH^i\left(\sqrt{\Lambda}\,  U_{i,\phi}(\phi)-H\,  W_{i}(\phi)\right)}{\sum_ii\,H^{i}  W_i(\phi)}, \\
\label{phi'}
 \phi'&=&\sqrt{\Lambda}\frac{\left(1-\Omega\right)H^2-3\sum_i(i-1)\,H^i\,  U_i(\phi)}{\sum_ii\,H^{i+1}  W_i(\phi)},
\end{eqnarray*}
where a prime represents derivative with respect to $\ln a$. We are now going to consider a number of cases in our search for viable cosmological models compatible with current observations. 

\subsection{Only $W_0 \neq 0$}

Let us first assume that $W_0 \neq 0$, and $W_1=W_2=W_3=0$. In this case $H'$ is ill defined. This can be understood by inspecting the Hamiltonian
\begin{eqnarray}
 \mathcal{H}_{\rm linear}&=&\sum_{i} \left[3(i-1)\sqrt{\Lambda}\,U_i(\phi)+i
\,\dot\phi\,W_i(\phi)\right]H^i \nonumber \\
&=& \sum_{i} \left[3(i-1)\sqrt{\Lambda}\,U_i(\phi)\right] H^i , \nonumber
\end{eqnarray}
which shows that it is independent of $\dot\phi$, therefore violating condition (ii) for a successful Lagrangian. This means that 
the model does not screen dynamically and  only the de Sitter solution exists.

\subsection{Only a $W_i$, $U_j$ pair}

From the constraint equation we have that $W_i = U_{j,\phi} \Lambda^{(j-i)/2}$, and then 
\begin{equation}
\frac{H'}{H} = -\frac{3}{i} \left[ 1-  \left( \frac{H}{\sqrt{\Lambda}} \right)^{j-i-1} \right], \nonumber
\end{equation}
which does not depend on $\phi$ and, consequently, the matter content has no influence on the Universe's evolution. When $j-i-1<0$ and $H \gg \sqrt{\Lambda}$, we can approximate the field equation by 
\begin{equation}
\frac{H'}{H} = -\frac{3}{i} . \nonumber
\end{equation}
We can, however,  obtain a dust like behaviour for $i=2$ and, as we expected by construction, we reach a de Sitter evolution when $H \rightarrow \sqrt{\Lambda}$.

\subsection{Only a $W_i$, $W_j$ pair}

In this case, from the constraint equation,  $W_i = -W_{j,\phi} \Lambda^{(j-i)/2}$ and then 
\begin{equation}
\frac{H'}{H} = -3 \frac{1-(H/\sqrt{\Lambda})^{i-j}}{j-i(H/\sqrt{\Lambda})^{i-j}} , \nonumber
\end{equation}
which is again independent of $\phi$. 
For $j>i$ and $H \gg \sqrt{\Lambda}$, the field equation reads,
\begin{equation}
\frac{H'}{H} = -\frac{3}{j} ,\nonumber
\end{equation}
and  we recover dust for $j=2$.  A de Sitter universe is reached when $H \rightarrow \sqrt{\Lambda}$.

\subsection{Term-by-Term model}

We now consider that the constraint equation is satisfied for equal powers of $\Lambda$, i.e. $W_i = 
U_{i,\phi}$. We have then eight functions and  four constraints, hence,  only four free potentials. 
Defining $U_{i,\phi} = \Lambda^{-i/2} V_{i,\phi}$, we can write,
\begin{equation}
\frac{H'}{H} = -3 \left( 1- \frac{\sqrt{\Lambda}}{H} \right)  \frac{\sum_i (H/
\sqrt{\Lambda})^i V_{i,\phi}}{\sum_i i (H/\sqrt{\Lambda})^i V_{i,\phi}}.\nonumber
\end{equation}
Here the field (and the background matter) contributes to the dynamics of
 the Universe as there is a dependence on $\phi$, which is determined by the matter content via Eq.~(\ref{phi'}). 
For $H\gg \sqrt{\Lambda}$ and when only one $i$ component dominates 
\begin{equation}
\frac{H'}{H} = -\frac{3}{i},\nonumber
\end{equation}
which means that we recover dust for $i=2$. As before, we reach de Sitter when $H \rightarrow \sqrt{\Lambda}$.

\subsection{Tripod model}

\begin{figure}
\begin{center}
\includegraphics[angle=0,width = 10cm]{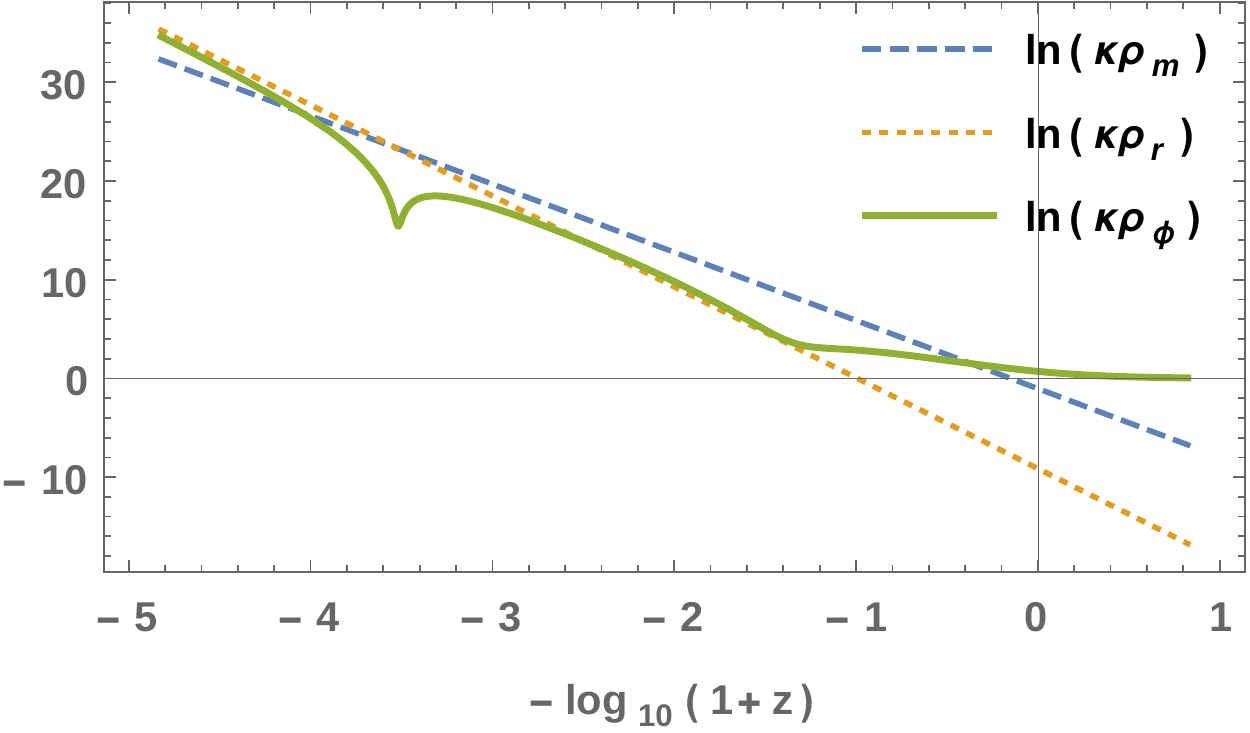} 
\caption{\label{tripod1} Energy densities evolution for the tripod models.}
\end{center}
\end{figure}

Let us consider the three potentials $U_2$, $U_3$ and $W_2$. The constraint equation
 imposes $U_{2,\phi} \Lambda + U_{3,\phi} \Lambda^{3/2} = W_2 \Lambda$, and then 
\begin{equation}
\frac{H'}{H} = -3 \frac{U_{2,\phi}}{W_2} \left(1- \frac{\sqrt{\Lambda}}{H} \right).\nonumber
\end{equation}
For $H\gg \sqrt{\Lambda}$, \hspace{0.5cm } 
\begin{equation}
\frac{H'}{H} = -\frac{3}{2}\frac{U_{2,\phi}}{W_2}, \nonumber
\end{equation}
therefore we need:
$U_{2,\phi}/W_2 = 1$, during a matter domination epoch, and 
$U_{2,\phi}/W_2 = 4/3$, for a radiation domination epoch.
For example, the choice of the potentials, $U_2 = e^{\lambda \phi} + \frac{4}{3} e^{\beta \phi}$, and
$W_2 = \lambda e^{\lambda \phi} + \beta e^{\beta \phi}$, give us the desired behaviour, as shown in Fig.~\ref{tripod1}.
The de Sitter evolution is attained when $H \rightarrow \sqrt{\Lambda}$.
Unfortunately, the contribution of the field at early times is too large to satisfy current constraints.

\section{Non-linear models}
In this section we consider the non-linear terms in $\dot\phi$ in the Lagrangian, i.e., 
\begin{equation}
 L_{\rm nl}=a^3 \sum_{i=0}^{3}X_i(\phi,\dot\phi)H^i, 
 \end{equation}
provided that any non-linear dependence of the Lagrangian on $\dot \phi$ vanishes at the 
critical point, thus, $\sum_{i=0}^{3} X_i(\phi,\dot\phi)\Lambda^{i/2}=0$, and combine it with $L_{\rm EH}$ and $L_{\rm m}$.
Again, $X_i$ are related to the the $\kappa_j$ functions of the Horndeski Lagrangian 
and $G_j$ functions of the Deffayet {\it et al.}~Lagrangian.
We will restrict the analysis to the  shift-symmetric cases, which means no dependence on $\phi$, and make use of the  convenient redefinition $\psi = \dot\phi$. Under these assumptions we obtain the equations of motion, 
\begin{eqnarray}
H'&=&\frac{3(1+w)Q_0P_1-Q_1P_0}{Q_1P_2-Q_2P_1}, \nonumber\\
\psi'&=&\frac{3(1+w)Q_0P_2-Q_2P_0}{Q_2P_1-Q_1P_2},\nonumber
 \end{eqnarray}
where $Q_0$, $Q_1$, $Q_2$, $P_0$, $P_1$, $P_2$, are complicated functions of $X_i$ and $H$, 
and the average equation of state parameter of matter fluids is 
\begin{equation}
1+w=\frac{\sum_s\Omega_{s}(1+w_s)}{\sum_s\Omega_{s}}.\nonumber
 \end{equation}
 As for the linear models, we are now going to take a systematic evaluation of the possible cosmological scenarios. 
 
 \subsection{$X_3 = \psi^n$ is the dominant contribution}

When $X_3$ is the dominant potential and $H\gg\sqrt{\Lambda}$, then the effective equation of state is
\begin{eqnarray}
1+w_{\rm eff}&\simeq&\frac{2}{3}(1+w),\qquad{\rm for}\qquad \frac{|\left(2 X_3+\psi X_{3,\psi}\right)X_{3,\psi\psi}|}{|\left(3X_{3,\psi}+\psi X_{3,\psi\psi}\right)X_{3,\psi}|} \gg 1 \nonumber \\
1+w_{\rm eff} &\simeq&\frac{2}{3}  \qquad{\rm otherwise}. \nonumber
\end{eqnarray}
Neither of these allow for $w_{\rm eff}$ corresponding to a radiation and/or matter domination epochs.

\subsection{$X_2 = \psi^n$ is the dominant contribution}

If instead $X_2$ is the dominant potential, for $H\gg\sqrt{\Lambda}$, it follows that  
\begin{eqnarray}
   w_{\rm eff} &\simeq& w,\qquad{\rm for}\qquad \frac{|\left(1-X_2-\psi X_{2,\psi}\right)X_{2,\psi\psi}|}{|\left(2X_{2,\psi}+\psi X_{2,\psi\psi}\right)X_{2,\psi}|} \gg 1, \nonumber \\ 
   w_{\rm eff} &\simeq& 0,\qquad{\rm otherwise}. \nonumber
\end{eqnarray}
In this case, either $w_{\rm eff}$ is too small today when compared with observational limits or, 
$\Omega_\psi$ is too large in the early Universe.

 \subsection{$X_0$ and $X_1$ are the sole contributions}

If we take $X_0$ and $X_1$ to be the only non-negligeble potentials, then it can be shown that when $H\gg\sqrt{\Lambda}$, the equation of state parameter  $w_{\rm eff}\simeq w$. This represents an
interesting case but, unfortunately, models with realistic initial conditions are not driven to the critical point.

\subsection{Extension with $X_0$, $X_1$ and $X_2$}

\begin{figure}
\begin{center}
\includegraphics[width=0.7\textwidth]{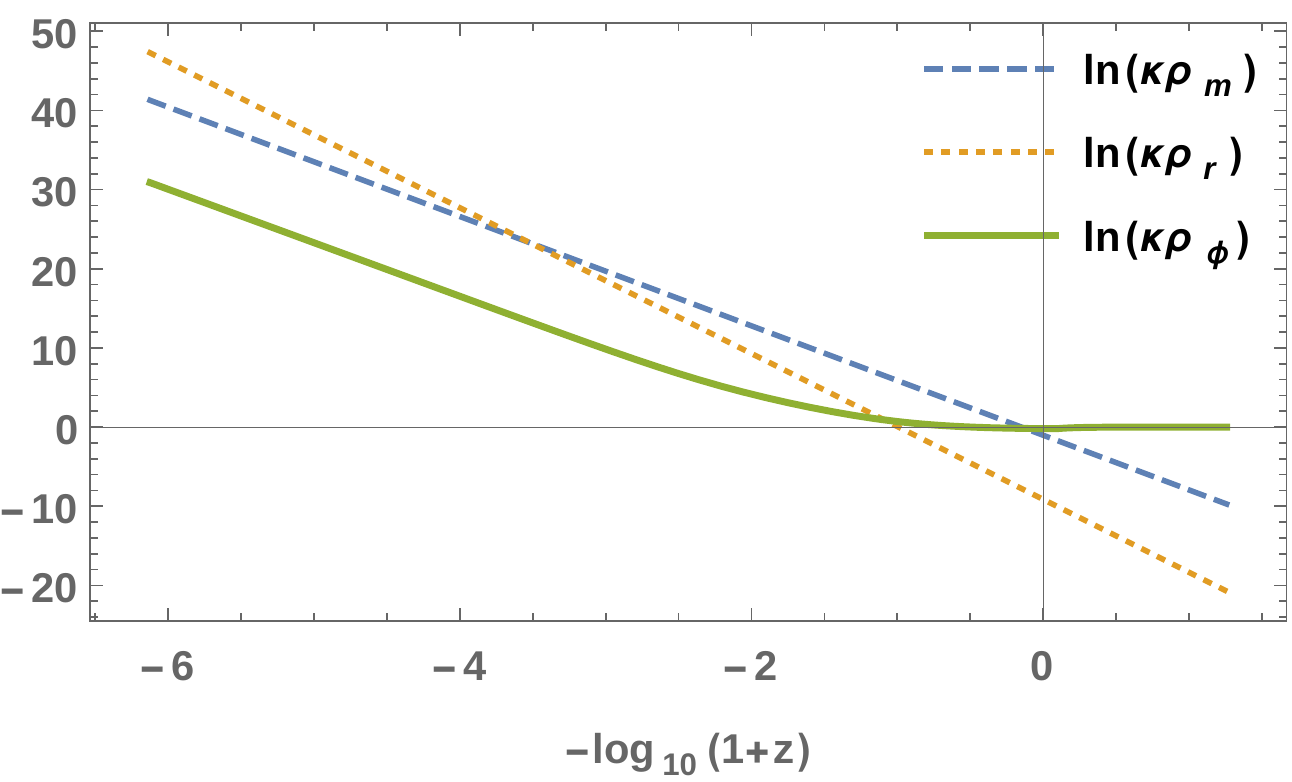}
\caption{\label{x0x1x2} Energy densities evolution for the model with non-vanishing $X_0$, $X_1$ and $X_2$. }
\end{center}
\end{figure}

Finally we consider a case involving the three potentials $X_0$, $X_1$ and $X_2$ such that
\begin{equation} 
X_2(\psi)=\alpha\psi^n,\qquad X_1(\psi)=-\alpha\psi^n+\frac{\beta}{\psi^m},\qquad X_0(\psi)=-\frac{\beta}{\psi^m}.\nonumber
\end{equation}
We can obtain a model with $w_\psi = w_0 + w_a (1-a)$, such that, $w_0 = -0.98$ and $w_a = 0.04$, which is compatible with current limits and moreover, has a negligible dark energy contribution at early times. The evolution of the energy densities is illustrated in Fig.~\ref{x0x1x2}.
 
 \section{Summary}
 
 In this article we have consider Horndeski cosmological models that may alleviate the
cosmological constant problem by screening any value of the vacuum energy given by
the theory of particle physics. We have presented the linear and the non-linear models and shown that the class of 
non-linear models with shift symmetry are in a better footing when they are compared with current observational constraints of the effective equation of state parameter and limits on early dark energy contribution. In order to further scrutinise these models, we are now required to face them against observables that depend on the evolution of the field and matter fluid fluctuations.


\section*{Acknowledgments}
This work was partially supported by the Funda\c{c}\~{a}o para a Ci\^{e}ncia e Tecnologia (FCT) through the grants EXPL/FIS-AST/1608/2013 and UID/FIS/04434/2013. PMM acknowledges financial support from the Spanish Ministry of Economy and Competitiveness (MINECO) through the postdoctoral training contract FPDI-2013-16161, and the project FIS2014-52837-P. FSNL was supported by a FCT Research contract, with reference IF/00859/2012.

\end{document}